\documentclass[aps,prl,groupedaddress,superscriptaddress,twocolumn,notitlepage,bibnotes]{revtex4-1}

\usepackage[latin1]{inputenc}
\usepackage{amsthm}
\usepackage{amssymb}
\usepackage{amsmath}
\usepackage{mathrsfs}
\usepackage{bbold}
\usepackage{bbm}
\usepackage{nonfloat}
\usepackage[pdftex]{hyperref}
\usepackage{braket}
\usepackage{dsfont}
\usepackage{mathdots}
\usepackage{mathtools}
\usepackage{enumerate}
\usepackage[shortlabels]{enumitem}
\usepackage{csquotes}
\usepackage{stmaryrd}
\usepackage[cal=boondox]{mathalfa}
\usepackage{graphicx}
\usepackage{stackengine}
\usepackage{scalerel}
\usepackage{array}
\usepackage{makecell}
\newcolumntype{x}[1]{>{\centering\arraybackslash}p{#1}}
\usepackage{tikz}
\usepackage{pgfplots}
\usepackage{ragged2e}
\usetikzlibrary{shapes.geometric, shapes.misc, positioning, arrows, arrows.meta, decorations.pathreplacing, decorations.pathmorphing, patterns, angles, quotes, calc}
\usepackage{booktabs}
\usepackage{xfrac}
\usepackage{siunitx}
\usepackage{centernot}
\usepackage{comment}
\usepackage{chngcntr}
\usepackage{caption}
\usepackage{subcaption}
\usepackage[normalem]{ulem}

\newtheorem{thm}{Theorem}
\newtheorem*{thm*}{Theorem}

\newtheorem*{prop*}{Proposition}

\newtheorem*{lemma*}{Lemma}

\newtheorem*{cor*}{Corollary}

\newtheorem*{cj*}{Conjecture}

\newtheorem*{Def*}{Definition}

\makeatletter
\def\thmhead@plain#1#2#3{%
  \thmname{#1}\thmnumber{\@ifnotempty{#1}{ }\@upn{#2}}%
  \thmnote{ {\the\thm@notefont#3}}}
\let\thmhead\thmhead@plain
\makeatother

\theoremstyle{definition}
\newtheorem{rem}[thm]{Remark}

\newtheorem{ex}[thm]{Example}

\newtheorem{manualthminner}{Theorem}
\newenvironment{manualthm}[1]{%
  \renewcommand\themanualthminner{#1}%
  \manualthminner \it
}{\endmanualthminner}

\newcommand{\bb}{\begin{equation}\begin{aligned}\hspace{0pt}}
\newcommand{\bbb}{\begin{equation*}\begin{aligned}}
\newcommand{\ee}{\end{aligned}\end{equation}}
\newcommand{\eee}{\end{aligned}\end{equation*}}
\newcommand*{\coloneqq}{\mathrel{\vcenter{\baselineskip0.5ex \lineskiplimit0pt \hbox{\scriptsize.}\hbox{\scriptsize.}}} =}
\newcommand*{\eqqcolon}{= \mathrel{\vcenter{\baselineskip0.5ex \lineskiplimit0pt \hbox{\scriptsize.}\hbox{\scriptsize.}}}}

\newcommand{\tmin}{\ensuremath \! \raisebox{2.5pt}{$\underset{\begin{array}{c} \vspace{-4.1ex} \\ \text{\scriptsize min} \end{array}}{\otimes}$}\!}
\newcommand{\tmax}{\ensuremath \! \raisebox{2.5pt}{$\underset{\begin{array}{c} \vspace{-4.1ex} \\ \text{\scriptsize max} \end{array}}{\otimes}$}\! }
\newcommand{\tminit}{\ensuremath \! \raisebox{2.5pt}{$\underset{\begin{array}{c} \vspace{-3.9ex} \\ \text{\scriptsize \emph{min}} \end{array}}{\otimes}$}\!}
\newcommand{\tmaxit}{\ensuremath \!\raisebox{2.5pt}{$\underset{\begin{array}{c} \vspace{-3.9ex} \\ \text{\scriptsize \emph{max}} \end{array}}{\otimes}$}\!}

\newcommand{\R}{\mathds{R}}

\newcommand{\C}{\mathds{C}}

\DeclareMathOperator{\co}{conv}
\DeclareMathOperator{\cone}{cone}

\DeclareMathAlphabet{\pazocal}{OMS}{zplm}{m}{n}

\DeclareMathOperator{\Id}{id}

\newcommand{\kite}{\mathsf{Q}}
\newcommand{\bsquare}{\mathsf{S}_b}

\newcommand{\lsmatrix}{\left(\begin{smallmatrix}}
\newcommand{\rsmatrix}{\end{smallmatrix}\right)}

\stackMath

\stackMath

\makeatletter
\newcommand*\rel@kern[1]{\kern#1\dimexpr\macc@kerna}
\newcommand*\widebar[1]{%
  \begingroup
  \def\mathaccent##1##2{%
    \rel@kern{0.8}%
    \overline{\rel@kern{-0.8}\macc@nucleus\rel@kern{0.2}}%
    \rel@kern{-0.2}%
  }%
  \macc@depth\@ne
  \let\math@bgroup\@empty \let\math@egroup\macc@set@skewchar
  \mathsurround\z@ \frozen@everymath{\mathgroup\macc@group\relax}%
  \macc@set@skewchar\relax
  \let\mathaccentV\macc@nested@a
  \macc@nested@a\relax111{#1}%
  \endgroup
}

\newcommand{\fakepart}[1]{
 \par\refstepcounter{part}
  \sectionmark{#1}
}

\counterwithin*{equation}{part}
\counterwithin*{thm}{part}
\counterwithin*{figure}{part}

\tikzset{meter/.append style={draw, inner sep=10, rectangle, font=\vphantom{A}, minimum width=30, line width=.8, path picture={\draw[black] ([shift={(.1,.3)}]path picture bounding box.south west) to[bend left=50] ([shift={(-.1,.3)}]path picture bounding box.south east);\draw[black,-latex] ([shift={(0,.1)}]path picture bounding box.south) -- ([shift={(.3,-.1)}]path picture bounding box.north);}}}
\tikzset{roundnode/.append style={circle, draw=black, fill=gray!20, thick, minimum size=10mm}}
\tikzset{squarenode/.style={rectangle, draw=black, fill=none, thick, minimum size=10mm}}

\definecolor{Blues5seq1}{RGB}{239,243,255}
\definecolor{Blues5seq2}{RGB}{189,215,231}
\definecolor{Blues5seq3}{RGB}{107,174,214}
\definecolor{Blues5seq4}{RGB}{49,130,189}
\definecolor{Blues5seq5}{RGB}{8,81,156}

\definecolor{Greens5seq1}{RGB}{237,248,233}
\definecolor{Greens5seq2}{RGB}{186,228,179}
\definecolor{Greens5seq3}{RGB}{116,196,118}
\definecolor{Greens5seq4}{RGB}{49,163,84}
\definecolor{Greens5seq5}{RGB}{0,109,44}

\definecolor{Reds5seq1}{RGB}{254,229,217}
\definecolor{Reds5seq2}{RGB}{252,174,145}
\definecolor{Reds5seq3}{RGB}{251,106,74}
\definecolor{Reds5seq4}{RGB}{222,45,38}
\definecolor{Reds5seq5}{RGB}{165,15,21}

\counterwithin*{equation}{part}
\counterwithin*{thm}{part}
\counterwithin*{figure}{part}

\tikzstyle{rounded1} = [rectangle, rounded corners, minimum width=2cm, minimum height=1cm,text centered, draw=black, fill=gray!20]
\tikzstyle{rounded2} = [rectangle, rounded corners, minimum width=2cm, minimum height=1cm,text centered, draw=black, fill=green!80!black]
\tikzstyle{arrow1} = [<->, very thick, green!50!black, >=stealth, line width=1mm]
\tikzstyle{arrow2} = [->, very thick, green!50!black, >=stealth, line width=1mm]

\newcommand{\Rp}{\mathds{R}_+}
\newcommand{\iy}{\infty}
\newcommand{\st}{\ : \ }
\renewcommand{\C}{C}
\newcommand{\CC}{\pazocal{C}}

\begin{document}

\title{Entanglement and superposition are equivalent concepts in any physical theory}

\author{Guillaume Aubrun}
\email{aubrun@math.univ-lyon1.fr}
\affiliation{Institut Camille Jordan, Universit\'e Claude Bernard Lyon 1, 43 boulevard du 11 novembre 1918, 69622 Villeurbanne CEDEX, France}

\author{Ludovico Lami}
\email{ludovico.lami@gmail.com}
\affiliation{Institute of Theoretical Physics and IQST, Universit\"{a}t Ulm, Albert-Einstein-Allee 11D-89069 Ulm, Germany}

\author{Carlos Palazuelos}
\email{carlospalazuelos@mat.ucm.es}
\affiliation{Departamento de An\'alisis Matem\'atico y Matem\'atica Aplicada, Universidad Complutense de Madrid, Plaza de Ciencias
s/n 28040 Madrid, Spain,}
\affiliation{Instituto de Ciencias Matem\'aticas, C/ Nicol\'as Cabrera, 13-15, 28049
Madrid, Spain}

\author{Martin Pl\'avala}
\email{martin.plavala@uni-siegen.de}
\address{Naturwissenschaftlich-Technische  Fakult\"{a}t, Universit\"{a}t Siegen, 57068 Siegen, Germany}

\begin{abstract}
We prove that any two general probabilistic theories (GPTs) are entangleable, in the sense that their composite exhibits either entangled states or entangled measurements, if and only if they are both non-classical, meaning that neither of the state spaces is a simplex. This establishes the universal equivalence of the (local) superposition principle and the existence of global entanglement, valid in a fully theory-independent way. As an application of our techniques, we show that all non-classical GPTs exhibit a strong form of incompatibility of states and measurements, and use this to construct a version of the BB84 protocol that works in any non-classical GPT.
\end{abstract}

\maketitle
\fakepart{Main text}

\textbf{\em Introduction.}---
When one looks back at the magnificent conceptual and philosophical revolution that quantum mechanics has sparked almost a century ago, two discoveries stand out as fraught with consequences, namely, the superposition principle and the existence of entanglement. The former entails that the behaviour of quantum systems cannot be described by classical probability theory, while the latter implies, via Bell's theorem~\cite{Bell, Brunner-review}, that the correlations exhibited by separate systems cannot be explained by means of local hidden variable models. These consequences of superposition and entanglement are predicted by the formalism of quantum mechanics, but they can be understood operationally, as simple statements concerning the frequencies of certain measurement outcomes. They can thus be regarded as theory-independent: any future `ultimate' theory of Nature, which may overcome quantum mechanics, must nevertheless encompass them and explain those experiments.

What is instead theory-\emph{dependent}, here, is the \emph{connection} between these two notions. Namely, it is only within the formalism of quantum theory that we can understand entanglement as the superposition principle applied to different product vectors of a tensor product Hilbert space~\cite{EPR, Horodecki-review}. That the connection between two fundamental phenomena whose physical existence rests on solid experimental evidence can only be understood by means of the mathematical formalism pertaining to a specific framework is somewhat conceptually unsatisfying, and, what is more, makes our understanding of said connection less sound and more dependent on the current theoretical paradigm --- which is, most likely, incomplete. And indeed, recently there have been several attempts to fill this gap and investigate the 
interplay between these two notions in an a priori fashion~\cite{Oppenheim-Wehner, Richens2017, Jencova2017, DAriano2019}.

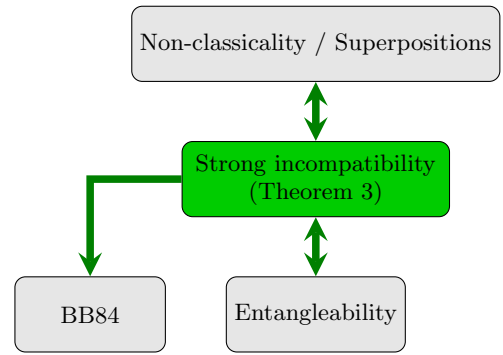
\begin{figure}
\begin{tikzpicture}
\node (nc) [rounded1] {Non-classicality / Superpositions};
\node (si) [rounded2, below of=nc, yshift=-0.8cm] {\begin{tabular}{c} Strong incompatibility \\ (Theorem~\ref{strong_incompatibility_thm}) \end{tabular}};
\node (ent) [rounded1, below of=si, yshift=-0.8cm] {Entangleability};
\node (bb84) [rounded1, left of=ent, xshift=-2cm] {BB84};
\node (aux) [left of=si] {};
\draw [arrow1] (nc) -- (si);
\draw [arrow1] (si) -- (ent);
\draw [arrow2] (si) -- ($(si)-(3cm,0)$) -- (bb84);
\end{tikzpicture}
\caption{Our path to connect the notions of non-classicality/superpositions and entangleability via that of strong incompatibility, which is also key to implementing the BB84 protocol.}
\label{conceptual_fig}
\end{figure}

In this paper we complete this programme, showing that the connection between superposition principle and entanglement --- and, moreover, a strong notion of incompatibility of measurements and states --- can be understood in a fully theory-independent way, and thus promoting it from a mere accident of the mathematics underpinning quantum mechanics to a logical necessity. This connection is demonstrated in Figure \ref{conceptual_fig}, showing that strong incompatiblity as given by Theorem \ref{strong_incompatibility_thm} connects non-classicality/superpositions with entangleability, at the same time allowing us to construct a version of the Bennett--Brassard 1984 (BB84) protocol~\cite{bennett1984quantum} in any non-classical theory.

In order to do this, we need a framework capable of encompassing all physical theories obeying minimal operational requirements, beyond standard quantum theory. The formalism of general probabilistic theories provides us with the widest possible arena to pursue such a programme~\cite{FOUNDATIONS, LUDWIG, Ludwig-1, Ludwig-2, Ludwig-3, Davies-1970}. A brief introduction can be found below; for a more complete one, we refer the reader to Ref.~\cite{lamiatesi, mueller2020, Plavala-intro}. Before explaining our result, we need to answer two questions.

(I) \emph{What does it mean that a certain state space exhibits superpositions?} The answer we shall adopt is that such state space should be \emph{non-classical}, i.e.\ it should not be described by a classical probability theory: that is, there should not be a finite set of `elementary states' that are both (a)~perfectly distinguishable by a measurement; and (b)~such that any other state can be written as a statistical mixture of them. In mathematical terms, this is equivalent to saying that the state space is not shaped as a simplex, the multi-dimensional generalisation of the two-dimensional triangle and of the three-dimensional tetrahedron. The connection we have made here between the existence of superpositions and the notion of non-classicality is a posteriori justified by Theorem~\ref{strong_incompatibility_thm} below, which illustrates it by drawing a striking parallel with quantum theory.

(II) \emph{What does it mean that two systems, i.e.\ two state spaces, exhibit entanglement?} First, we need to distinguish between entanglement at the level of states and entanglement at the level of measurements. The former means that there are states on the bipartite system that cannot be written as a statistical mixture of uncorrelated (product) states. The latter, accordingly, means that there are bipartite effects that are not a positive linear combination of product effects.

\textbf{\em General probabilistic theories.}---
In the most general sense, a physical theory is simply a set of rules that allow to deduce a probabilistic prediction of the outcome of an experiment given the detailed description of its preparation. From this abstract description one can deduce, via the so-called Ludwig's embedding theorem~\cite{Ludwig-1, LUDWIG, lamiatesi}, the mathematical formalism of general probabilistic theories (GPTs) that we will now describe~\cite{lamiatesi, mueller2020, Plavala-intro}.

The fundamental object needed to model an arbitrary physical system is its \emph{state space}; mathematically, this will be represented as a generic convex and compact subset $\Omega$ of some finite-dimensional real vector space~\footnote{The assumption of finite dimension is a technical one. It is possible and in general desirable to drop it, although that comes at the cost of significantly increasing the mathematical complexity of the theory~\cite[Chapter~1]{lamiatesi}.}. Physically, a state $\omega\in \Omega$ should be thought of as a description of a preparation procedure for the system under examination. The convexity of $\Omega$ reflects the fact that preparation procedures can be mixed stochastically: the ensemble $\{ p_i, \omega_i\}$, which corresponds to the physical procedure of drawing a random variable $I$ and preparing the system in the state $\omega_i$, is represented within the formalism by the convex mixture $\sum_i p_i \omega_i$.

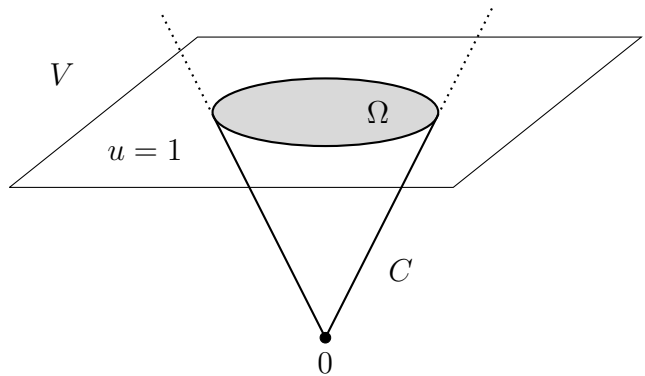
\begin{figure}
\begin{tikzpicture}[scale=1]
\draw[fill=gray!30,thick] (0,3) circle [x radius=1.5cm, y radius=0.45cm];
\coordinate (origin) at (0,0);
\filldraw (0,0) circle (2pt);
\node at (0,-0.33) {\large $0$};
\draw[thick] (origin) -- (-1.491,2.95);
\draw[thick] (origin) -- (1.491,2.95);
\draw[thick,dotted] (1.491,2.95) -- (2.2365, 4.425);
\draw[thick,dotted] (-1.491,2.95) -- (-2.19177, 4.3365);
\draw (-4.2,2) -- (1.7,2) -- (4.2,4) -- (-1.7,4) -- (-4.2,2);
\node (Omega) at (0.7,3) {\large $\Omega$};
\node (C) at (1,0.9) {\large $C$};
\node (u) at (-2.4,2.5) {\large $u=1$};
\node (V) at (-3.5,3.5) {\large $V$};
\end{tikzpicture}
\caption{The basic ingredients of a GPT are a real finite-dimensional vector space $V$ and a cone $C$. The order unit functional $u$ defines a hyperplane $u^{-1}(1)$, whose intersection with $C$ identifies the state space $\Omega$.}
\label{cone with section}
\end{figure}

It is often useful to include into the picture not only normalised but also un-normalised states. This can be done by imagining a situation as the one depicted in Figure~\ref{cone with section}. Namely, in an augmented vector space $V$ we introduce a proper cone $C$, i.e.\ a set $C\subset V$ that is closed under positive scalar multiplication, and moreover: (i)~convex; (ii)~salient, meaning that $C\cap (-C)=\{0\}$; (iii)~generating, in the sense that $C-C=V$; and (iv)~topologically closed. The state space $\Omega$ is then recovered as the section of $C$ identified by the equation $u=1$, where $u\in V^*$ is a `normalising' functional, called the \emph{order unit}, belonging to the dual vector space $V^*$ and (v)~strictly positive on $C$, i.e.\ such that $u(x)> 0$ for all $x\in C$ with $x\neq 0$. We can summarise the above discussion by giving an abstract definition of a GPT as any triple $(V,C,u)$, where $V$ is any real finite-dimensional vector space, $C\subset V$ is a proper cone inside it, and $u\in V^*$ is a strictly positive functional on $C$.

From the mathematical standpoint, the introduction of the (proper) cone $C$ makes $V$ an ordered vector space: for any two $x,y\in V$, we define the ordering by stipulating that $x\leq y$ if $y-x\in C$. Notably, this ordering is not total, i.e.\ it is possible that neither $x\leq y$ nor $y\leq x$. The dual space $V^*$ inherits an ordering from $V$: for $f,g\in V^*$, we write $f\leq g$ if $f(\omega)\leq g(\omega)$ for all $\omega\in \Omega$ (equivalently, $f(x)\leq g(x)$ for all $x \in C$). The cone of positive functionals in $V^*$, called the \emph{dual cone} to $C$, is denoted with $C^*$. Remarkably, for proper cones $C$ we have the identity $C^{**} = C$.

To complete our picture we need to discuss measurements alongside with states. The description of a physical measurement together with the identification of one of its possible outcomes will be represented mathematically by an \emph{effect}. This is just a linear functional $e\in V^*$; the value $e(\omega)$ it takes on a state $\omega\in \Omega$, which must be comprised between $0$ and $1$, is interpreted as the probability that the corresponding outcome occurs when that state is measured. Employing the above notion of ordering on $V^*$, we can compactly require that $0\leq e\leq u$. A fully-fledged measurement will then be a (finite) collection of effects $(e_i)_{i\in I}$, where $e_i\in V^*$ with $e_i\geq 0$ for all $i$. The normalisation condition for the outcome probabilities implies that $\sum_{i\in I} e_i = u$.

Now that we have a mathematical description of measurements, we could wonder whether such description is complete. Namely, given a collection of effects summing to $u$, can it be physically implemented as a measurement procedure? If that is the case, we will say that the system satisfies the \emph{no-restriction hypothesis}~\cite{no-restriction, Chiribella-pur}. We deem this a natural assumption, for the good reason that classical theories (Example \ref{example classical}) and quantum mechanics satisfy it. Thus, throughout this paper we will always include the no-restriction hypothesis in our theoretical framework.

The GPT formalism we have just sketched may appear rather abstract. To make it more concrete, let us discuss a very important case. 

\begin{ex}[(Classical theories as GPTs)] \label{example classical}
In classical theory the cone $C$ is generated by a set of linearly independent states. It then follows that the state space is a simplex and that every state is given as a unique convex combination of the generating states.
\end{ex}

We can now give a precise mathematical meaning to the answer to question~(I) in the Introduction. Within the formalism of GPTs, we identify the existence of abstract superpositions with the non-classicality of the theory, where we say that a GPT is \emph{non-classical} if it is not of the form described in Example~\ref{example classical}.

\textbf{\em Bipartite systems.}---
In order to describe entanglement we need to introduce bipartite systems into the picture. Given two systems $A,B$ modelled by GPTs $A=(V_1, C_1, u_1)$ and $B=(V_2, C_2, u_2)$, can we represent also the joint system $AB$ as a GPT $AB=\left(V_{12}, C_{12}, u_{12}\right)$? In this context, a natural assumption --- which we shall adopt throughout the paper --- is the so-called \emph{local tomography principle}. In layman's terms, it states that the composite system should not contain more degrees of freedom than its parts. In more mathematical terms, we require that the statistics under product measurements determine any state of the bipartite system uniquely. With this assumption, one can prove the familiar tensor product rule~\cite{tensor-rule-1, tensor-rule-2}
\begin{equation}
V_{12} = V_1\otimes V_2\, ,\quad u_{12}=u_1\otimes u_2\, .
\label{tensor rule space}
\end{equation}
Two operationally motivated constraints on the cone $C_{12}$ come from the fact that independent local actions, namely, state preparations and measurements, should be faithfully represented in the bipartite picture as well. More formally, (i)~local (tensor product) states should also be valid bipartite states, and (ii)~local (tensor product) effects should also be valid effects on the bipartite system. Introducing the \emph{minimal} and the \emph{maximal tensor product} of the cones $C_1$ and $C_2$, defined by
\begin{align}
C_1\tmin C_2 &\coloneqq \co \left\{ x\otimes y:\ x\in C_1,\, y\in C_2\right\} , \label{C min} \\
C_1\tmax C_2 &\coloneqq \left( C_1^* \tmin C_2^* \right)^* , \label{C max}
\end{align}
where $\co$ denotes the convex hull, we can rephrase~(i) as $C_1\tmin C_2\subseteq C_{12}$ and~(ii) as $C_1^*\tmin C_2^*\subseteq C_{12}^*$. By combining the former relation with the dual of the latter we obtain the two-fold bound
\begin{equation}
C_1 \tmin C_2 \subseteq C_{12} \subseteq C_1 \tmax C_2
\label{double bound}
\end{equation}
on the bipartite cone $C_{12}$. (Note that by definition $C_1 \tmin C_2 \subseteq C_1 \tmax C_2$.) We are now in the position to formalise the answer to question~(II) in the Introduction: the existence of entanglement at the level of states or at the level of measurements is equivalent to one of the two inclusions in~\eqref{double bound} being strict. In turn, this happens if and only if
\bb
C_1\tmin C_2\neq C_1\tmax C_2\, ,
\label{entangleability}
\ee
i.e.\ if the minimal tensor product is a strict subset of the maximal tensor product. When this is the case we will say that the two GPTs $A,B$ are \emph{entangleable}. One interesting aspect of this definition of entangleability is that it does away with the need to specify the bipartite cone: whatever $C_{12}$ is chosen to be,~\eqref{entangleability} guarantees that the joint system will exhibit either entangled states or entangled measurements (or both).

\textbf{\em Entangleability.}---
Our result on entangleability, whose (highly technical) proof can be found in \cite{cones-2}, is as follows.

\begin{thm} \label{entangleability_thm}
Two GPTs $A,B$ are entangleable if and only if they are both non-classical.
\end{thm}

The above theorem pinpoints a profound and intrinsic connection between the notions of non-classicality and entanglement: the two concepts are two sides of the same coin, and not merely linked by a mathematical accident of the quantum mechanical formalism. Theorem~\ref{entangleability_thm} relies on two main assumptions: first, the no-restriction hypothesis, positing that every mathematically consistent effect is physically realisable; and second, the local tomography principle, which entails that combining two systems does not lead to the appearance of new degrees of freedom. These two assumptions are not only natural, because they are satisfied by both classical theories and quantum theory, but also necessary to avoid the mathematical trivialisation of the problem. In fact, by dropping the no-restriction hypothesis it is possible to enforce a minimal tensor product composition rule at the level of states and of measurements at the same time, eliminating entanglement somewhat artificially. On the other hand, without the local tomography principle the dimension of the linear span of $C_{12}$ is larger than that of the span of $C_1\tmin C_2$, directly implying the existence of entangled states~\cite[Proposition~2]{DAriano2019}. 

Note that one implication is easy: if either $A=(V_1,C_1,u_1)$ or $B=(V_2,C_2,u_2)$ is classical, then it is not difficult to see that $C_1\tmin C_2=C_1\tmax C_2$, meaning that $A$ and $B$ are not entangleable~\cite{Barker1976, Barker-review}. The converse implication is the truly challenging one.

As it is formulated now, Theorem~\ref{entangleability_thm} is rather abstract, as it merely asserts the existence of entangled objects (either states or measurements) in certain composite theories; it tells us nothing about how that entanglement may be detected and harnessed, and what it may be useful for. To remedy this, in what follows we will show that our result in fact leads to the construction of a BB84 protocol~\cite{bennett1984quantum} that allows for secret key distribution over a public noiseless channel in any non-classical GPT.

\textbf{\em Strong incompatibility.}---
We consider now a strengthened version of the well-known notion of incompatibility and prove that it is in fact fully equivalent to non-classicality (see Figure~\ref{conceptual_fig}). Given a vector space $V$ ordered by a cone $C$, two finite families of vectors
$x_i \in C$ and $y_j \in C$ are said to be compatible if one can find $z_{ij}\in C$, such that $\sum_j z_{ij} = x_i$ and $\sum_i z_{ij}=y_j$ for all $i,j$;
they are said to be incompatible otherwise. Clearly, a necessary but in general not sufficient condition for compatibility is that $\sum_i x_i = \sum_j y_j$. If $(V,C,u)$ forms a GPT, we can try to find incompatible vectors either in the primal space $V,C$ or in the dual space $V^*,C^*$. This latter case is particularly important operationally, as two measurements are compatible if and only if they can be implemented jointly~\cite{Heinosaari2016}.

The connection between incompatibility and non-classicality of GPTs has been explored thoroughly~\cite{Busch-1986, Wolf-incompatible, Busch-2013, Banik-2013, Stevens-Busch, Cavalcanti-2016, Plavala-2016, Jencova2017, Jencova-Plavala}. For instance, it is known that a GPT is non-classical if and only if it admits two incompatible binary measurements~\cite{Plavala-2016}. Here we establish a modified and somewhat stronger version of this fact:
\begin{thm} \label{strong_incompatibility_thm}
A proper cone $C$ is non-classical if and only if there are non-zero vectors $0\neq x_0,x_1,x_+,x_-\in C$ and functionals $f_0,f_1,f_+,f_-\in C^*$ such that:
\begin{enumerate}[(i)]
\item $x_0+x_1 = x_++x_-$ and $f_0+f_1=f_++f_-$;
\item $f_0(x_1) = f_1(x_0) = f_+(x_-) = f_-(x_+) = 0$;
\item $f_i+f_j$ is strictly positive, for all $i\in \{0,1\}$, $j\in \{+,-\}$.
\end{enumerate}
\end{thm}

The proof of Theorem \ref{strong_incompatibility_thm} can be found in the Supplemental Material~\footnote{See the SM, which contains Ref.~\cite{cones-1}, for further details.\label{SM}}. At first sight it may not be clear what Theorem~\ref{strong_incompatibility_thm} has to do with the notion of incompatibility. However, the two families of vectors $x_0, x_1$; $x_+,x_-$ constructed there are in fact incompatible. To see this, assume that a decomposition $(z_{ij})_{ij}\in C$ holds, so that $\sum_j z_{ij} = x_i$ and $\sum_i z_{ij} = x_j$. Then $0=f_1(x_0) = f_1(z_{0+}+z_{0-}) \geq f_1(z_{0+})\geq 0$ and analogously $0=f_-(x_+) = f_-(z_{0+}+z_{1+}) \geq f_-(z_{0+})\geq 0$, so that $f_1(z_{0+})=f_-(z_{0+})=0$. Since $f_1+f_-$ must be strictly positive and $(f_1+f_-)(z_{0+})=0$, it necessarily holds that $z_{0+}=0$. Repeating this reasoning we reach the absurd conclusion that $z_{ij}\equiv 0$ for all $i,j$; hence, the vectors $x_0, x_1$; $x_+,x_-$ were indeed incompatible. 

Note that Theorem \ref{strong_incompatibility_thm} supports the idea of identifying superposition with non-classicality (as we have done) since one can draw a direct parallel between the two families of vectors $x_0, x_1$; $x_+,x_-$ and the vectors $\ket{0}, \ket{1}$; $\ket{+}, \ket{-}$ representing states of a qubit, where $\ket{\pm} = \frac{1}{\sqrt{2}}(\ket{0} \pm \ket{1})$. The corresponding effects $f_0, f_1$; $f_+, f_-$ are then simply the projections onto the subspaces generated by the respective vectors. It is then clear that (ii)~implies that we can treat $x_0, x_1$ and $x_+,x_-$ as operational generalization of two different orthonormal sets in quantum theory, while (i)~implies that the linear hulls of these sets overlap. In this sense, one can also interpret the result of Theorem~\ref{strong_incompatibility_thm} as stating that any non-classical state space exhibits an operational form of quantum discord \cite{OllivierZurek-discord,AdessoCianciarusoBromley-discord}.

Theorem~\ref{strong_incompatibility_thm} allows us to provide a simple proof of Theorem \ref{entangleability_thm}~\cite{Note2}. Indeed, given two non-classical GPTs $A=(V_1,C_1,u_1)$ and $B=(V_2,C_2,u_2)$, thanks to Theorem~\ref{strong_incompatibility_thm} we can construct an explicit tensor belonging to $C_1\tmax C_2$ but not to $C_1\tmin C_2$, thus demonstrating~\eqref{entangleability}. In order to do this, we invoke Theorem~\ref{strong_incompatibility_thm} for the cone $C_1$ (resp., $C_2$) to construct vectors $0\neq x_0,x_1; x_+,x_- \in C_1$ and functionals $f_0,f_1; f_+,f_- \in C_1^*$ (resp., vectors $y_0,y_1; y_+,y_- \in C_2$ and functionals $g_0,g_1; g_+, g_- \in C_2^*$) satisfying conditions (i)--(iii). We then construct the state
\begin{equation} \label{eq:entangled-omega}
\omega = x_0 \otimes y_+ - x_+ \otimes y_+ + x_+ \otimes y_0 + x_1 \otimes y_1 \, .
\end{equation}
It turns out that
\begin{equation}
\omega \in C_1\tmax C_2 \setminus C_1\tmin C_2\, ,
\label{eq:omega-cones}
\end{equation}
thus implying~\eqref{entangleability}, i.e.\ $A,B$ are entangleable. The proof of~\eqref{eq:omega-cones} consists of two parts: first we show that $\omega \in C_1 \tmax C_2$, which is rather straightforward and follows from (i). To show that $\omega \notin C_1 \tmin C_2$ we construct a Bell-like inequality of the Clauser--Horne--Shimony--Holt (CHSH) type~\cite{CHSH} using the functionals $f_0,f_1; f_+,f_-$ and we prove that this inequality is violated. Note that since in general $f_0+f_1 = f_++f_- \neq u$, the aforementioned Bell inequality is not necessarily a Bell inequality in the underlying GPTs and the question whether any two non-classical GPTs violate some Bell inequality is still open.

\textbf{\em Application: BB84 protocol in GPTs.---}
As the main application of the theory developed here we show how to design a version of the BB84 protocol~\cite{bennett1984quantum} for secret key distribution over a public channel that works in any non-classical GPT. The motivation follows from the aforementioned parallel between the families of vectors $x_0, x_1; x_+, x_-$ and the vectors $\ket{0}, \ket{1}; \ket{+}, \ket{-}$. Since the later are used to construct the BB84 protocol in quantum theory, it is natural to ask whether the former allow us to do the same in any non-classical GPT.

The main idea is rather straightforward: let $(V,C,u)$ be a non-classical GPT and let and let $x_0,x_1$; $x_+,x_-$ and $f_0,f_1$; $f_+,f_-$ be the vectors as given by Theorem~\ref{strong_incompatibility_thm}. We can construct states $\rho_0, \rho_1; \sigma_+, \sigma_- \in \Omega$ such that $p_i \rho_i = x_i$ and $q_j \sigma_j = x_j$ for some $p_i, q_j >0$. It then follows that $p_0 \rho_0 + p_1 \rho_1 = q_+ \sigma_+ + q_- \sigma_-$. By re-scaling if necessary, we can assume that $p_0+p_1=q_++q_-=1$ and similarly that $f_0+f_1 = f_++f_- \eqqcolon \ell \leq u$.

Now, Alice tosses a fair coin; if heads, she prepares one of the states $\rho_0, \rho_1$ (with a priori probabilities $p_0,p_1$); if tails, one of the states $\sigma_+, \sigma_-$ (with a priori probabilities $q_+,q_-$). Since $p_0 \rho_0 + p_1 \rho_1 = q_+ \sigma_+ + q_- \sigma_-$, an eavesdropper Eve cannot discern these two scenarios. Unlike in the quantum case, it is not guaranteed that Bob can perfectly discriminate the ensembles $\rho_0,\rho_1$ or $\sigma_+,\sigma_-$; however, he will toss a fair coin too, and run an unambiguous state discrimination procedure using the measurements $f_0,f_1,u-\ell$ (if heads) or $f_+,f_-,u-\ell$ (if tails). This introduces an additional error, as the rounds where Bob obtains the outcome $u-\ell$ have to be discarded. Despite that, Alice and Bob can proceed in the usual way: they make the results of their coin tosses public and they remove the rounds for which either the choices of preparation and measurement were not the same or Bob obtained the outcome $u-\ell$. In the remaining cases the choices of preparation and measurement correspond, and moreover Bob's outcome was not $u-\ell$. Using Theorem~\ref{strong_incompatibility_thm}(ii), we thus see that Bob has recovered with no error the key bit $i$. In this way Alice and Bob obtain a shared key. One of the significant differences with the quantum case is that this key is not automatically secret. In fact, the information revealed to Eve \emph{is} correlated with the key bit. To remedy this, Alice and Bob can run the secret key distillation protocol proposed by Maurer~\cite{Maurer1993} to extract a truly secure key. A detailed description of the protocol as well as proof that our version of it achieves a non-zero secret key generation rate can be found in the Supplemental Material~\footnotemark[2].

\textbf{\em Conclusions.---}
We have showed that the connection between superpositions, entanglement, and BB84 protocol is purely operational and exists in every non-classical GPT. The crucial aspect of our techniques is that they by-pass the Hilbert space structure that underlies quantum mechanics, but that is not included in other possible non-classical theories. This gives a counter-example to possible axiomatizations of quantum theory \cite{Hardy2001}: for example, it is known that existence of purifications \cite{Chiribella-pur, Chiribella-info-der}, certain symmetries \cite{Masanes2011,Garner2017} or self-duality and spectrality \cite{Barnum2019} are enough to single-out quantum theory among other non-classical theories. Our results show that existence of superpositions, entanglement, and availability of BB84 protocol do not restrict the set of possible theories at all.

Our main method was to exploit the strong incompatibility inherent in every non-classical GPT. Strong incompatibility allowed us to construct the universal entangled tensor~\eqref{eq:entangled-omega}, but also a generalised version of the BB84 protocol that works in any non-classical operational theory. It is an open question whether one can derive other properties characterising non-classical GPTs, such as no-broadcasting \cite{Barnum-no-broad}, from strong incompatibility. It is also open whether the violations of Bell inequalities and steering exist in any non-classical GPT; we anticipate that some version of strong incompatibility may play an important role in investigating this question.

\begin{acknowledgments}
\textbf{\em Acknowledgments.---}
GA was supported in part by ANR (France) under the grants StoQ (2014-CE25-0003) and ESQuisses (ANR-20-CE47-0014-01). LL acknowledges financial support from the European Research Council under the Starting Grant GQCOP (Grant no.~637352), from the Foundational Questions Institute under the grant FQXi-RFP-IPW-1907, and from the Alexander von Humboldt Foundation. CP is partially supported by Spanish MINECO through Grant No.~MTM2017-88385-P, by the Comunidad de Madrid through grant QUITEMAD-CM P2018/TCS4342 and by SEV-2015-0554-16-3. MP acknowledges support from the Deutsche Forschungsgemeinschaft (DFG, German Research Foundation, project numbers 447948357 and 440958198), the Sino-German Center for Research Promotion (Project M-0294), the ERC (Consolidator Grant 683107/TempoQ), and from the Alexander von Humboldt Foundation.
\end{acknowledgments}

\bibliographystyle{apsrev4-1}
\bibliography{cones}

\clearpage
\fakepart{Supplemental Material}

\onecolumngrid
\begin{center}
\vspace*{\baselineskip}
{\textbf{\large Supplemental Material}}\\
\end{center}

\renewcommand{\theequation}{S\arabic{equation}}
\renewcommand{\thethm}{S\arabic{thm}}
\renewcommand{\thefigure}{S\arabic{figure}}
\setcounter{page}{1}
\makeatletter

\setcounter{secnumdepth}{2}

\section{Proofs of Theorem~\ref{entangleability_thm} and Theorem~\ref{strong_incompatibility_thm}}

In this part of the supplemental material we prove Theorem~\ref{strong_incompatibility_thm} in the main text. Leveraging this latter result, we will present a direct proof of Theorem~\ref{entangleability_thm} that is under several aspects more intuitive and physically and operationally meaningful than that reported in~\cite{cones-2}.
However, we warn the reader that the demonstration of Theorem~\ref{strong_incompatibility_thm} is based on a key result from \cite{cones-2} whose highly technical proof will not be reproduced here.

Let us first recall some basic notions on convex cones. Here, all the vector spaces are assumed to be real and finite-dimensional; they are denoted with symbols such as $V,V'$, and so on. A subset $\C$ of a vector space $V$ is called a \emph{cone} if it satisfies $sx+ty \in \C$ for every $x$, $y \in \C$ and $s$, $t \in \Rp$ ($\Rp$ stands for the half-line $[0,\iy)$ of non-negative reals). The cone generated by a subset $A \subset V$ will be denoted with $\cone(A)$.

A cone $\C \subset V$ is said to be \emph{generating} if $\C - \C = V$ (equivalently, if it spans $V$ as a vector space). Also, $\C$ is called \emph{salient} (or \emph{pointed}) if it does not contain a line, i.e.\ if $\C \cap (-\C) = \{0\}$. Finally, $\C$ is \emph{proper} if it is closed, salient and generating. 
We call \emph{convex body} a compact convex subset of a vector space with a nonempty interior. 

For a convex set $K \subset V$, we can consider the \emph{cone over $K$}, which is the cone in $V \times \R$ defined by
\[ \CC(K) = \cone( K \times \{1\} ) = \{ (x \,;\,t) \in V\times \Rp \st x \in tK \}. \]
If $K$ is a convex body, then $\CC(K)$ can be shown to be a proper cone.

Let $V$ be a vector space, and  $V^*$ its dual space, i.e.\ the space of linear functionals on $V$. Given a cone $\C$ in $V$, we can construct its \emph{dual cone}, defined as
\[ \C^* = \{ f \in V^* \st f(x) \geq 0  \textnormal{ for every }\,  x \in \C \} .\]
The bipolar theorem states that if $C$ is a closed cone then $\C = (\C^*)^*$, up to the canonical identification of $V$ with the bidual $V^{**}$.

Let $\C$ be a cone. An element $x \in \C$ is called an \emph{extreme ray generator} if $x=y+z$, for $y$, $z \in \C$, implies that $y=\alpha x$ for some $\alpha\in [0,1]$. If that is the case, the set $\{tx \st t\in\Rp\}$ is said to be an \emph{extreme ray} of $\C$. An element $f\in V^*$ is said to be \emph{strictly positive} if $f(x)>0$ for every $x\in C$ with $x\neq 0$.

\begin{figure}[h!] \begin{center}
\begin{tikzpicture}[scale=2]
	\coordinate (a) at (1,1);
	\coordinate (b) at (1,-1);
	\coordinate (c) at (-1,-1);
	\coordinate (d) at (-1,1) ;
	\coordinate (A) at (0.7,1);
	\coordinate (B) at (1,0.9);
	\coordinate (C) at (-0.2,-1);
	\coordinate (D) at (-1,0.5) ;
	\draw[fill=gray!30] (a)--(b)--(c)--(d)--(a) ;
	\draw[fill=gray!90] (A)--(B)--(C)--(D)--(A) ;
	\draw[gray!0] (a) node {$\bullet$};
        \draw[gray!0] (b) node {$\bullet$};
        \draw[gray!0] (c) node {$\bullet$};
        \draw[gray!0] (d) node {$\bullet$};
	\draw (-0.1,-0.1) node {$\kite_{\alpha}$};
	\draw (0.6,-0.5) node {$\bsquare$};
        \end{tikzpicture} \end{center}
\caption{A kite inside the blunt square}
\label{fig:kite-square}
\end{figure}
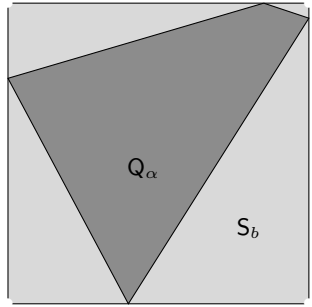

We need to define two particular planar convex shapes which will be crucial for us, see Figure \ref{fig:kite-square}. First, the \emph{blunt square} is constructed as the unit square minus its corners, via the formula
\[ \bsquare \coloneqq [-1,1]^2 \setminus \{-1,1\}^2. \]
Also, we define a \emph{kite} as
\begin{equation}
\kite_\alpha = \co \{ (-1,\alpha_0), (1,\alpha_1), (\alpha_-,-1), (\alpha_+,1) \} ,
\end{equation}
where $\alpha=(\alpha_0,\alpha_1,\alpha_-,\alpha_+) \in (-1,1)^4$. Note that any kite is a subset of the blunt square.

Let $\C$ be a proper cone in a finite-dimensional vector space $V$. Borrowing the terminology from~\cite{cones-2}, we will say that $\C$ \emph{admits a kite-square sandwiching} if one can find a kite $\kite_\alpha$ and two linear maps $\Psi : \R^3 \to V$, $\Phi : V \to \R^3$ with the property that $\Phi \circ \Psi = \Id$, $\Psi(\CC(\kite_\alpha)) \subset \C$ and moreover $\Phi(\C) \subset \CC(\bsquare)$. 

Then, a crucial result proved in \cite{cones-2} is the following.
\begin{thm} \label{theorem:KS}
A proper cone $\C$ is non-classical if and only if it admits a kite-square sandwiching.
\end{thm}

\begin{rem} \label{rem:ker-Phi}
The argument from \cite{cones-2} gives an extra piece of information of which we will make use: the map $\Phi : V \to \R^3$ involved in the kite-square sandwiching satisfies $C \cap \ker \Phi = \{0\}$.
\end{rem}

As we explain in the main text, the \emph{minimal} and the \emph{maximal tensor product} of the cones $\C_1$ and $\C_2$ are defined by
\begin{align}
\C_1\tmin \C_2 &\coloneqq \co \left\{ x\otimes y:\ x\in \C_1,\, y\in \C_2\right\}, \\
\C_1\tmax \C_2 &\coloneqq \left(\C_1^* \tmin \C_2^* \right)^*. 
\end{align}

It is very easy to show that $C_1\tmin C_2 \subseteq C_1 \tmax C_2$. The result on entangleability in the main text can be then stated in the following way.

\begin{manualthm}{\ref{entangleability_thm}}
Let $\C_1$ and $\C_2$ be a proper cones. Then, they are both non-classical if and only if
\begin{align*}
\C_1\tminit \C_2\neq \C_1\tmaxit \C_2\, .
\end{align*}
\end{manualthm}

This result was proved in \cite{cones-2} and its proof crucially relies on Theorem~\ref{theorem:KS}, which appears as a purely mathematical result with no clear physical interpretation. Here, we will first show how Theorem~\ref{strong_incompatibility_thm} in the main text can be obtained from Theorem~\ref{theorem:KS}. 
\begin{manualthm}{\ref{strong_incompatibility_thm}}

A proper cone $\C$ is non-classical if and only if there are non-zero vectors $0\neq x_0,x_1,x_+,x_-\in \C$ and functionals $f_0,f_1,f_+,f_-\in \C^*$ such that:
\begin{enumerate}[(i)]
\item $x_0+x_1 = x_++x_-$ and $f_0+f_1=f_++f_-$;
\item $f_0(x_1) = f_1(x_0) = f_+(x_-) = f_-(x_+) = 0$;
\item $f_i+f_j$ is strictly positive, for all $i\in \{0,1\}$, $j\in \{+,-\}$.
\end{enumerate}

\end{manualthm}
\begin{rem}\label{remark supp}
It is easy to see that the previous items imply that $f_0+f_1 = f_-+f_+$ is also strictly positive.
\end{rem}
\begin{proof}
It follows from the comments right after Theorem~\ref{strong_incompatibility_thm} in the main text that if $\C$ is a classical cone, there cannot exist elements $x_i$ and functionals $f_i$ satisfying properties (i)--(iii): since if $\C$ is classical, all elements are compatible. Hence, we just need to prove the converse implication.

To this end, let us assume that $\C$ is non-classical. According to Theorem \ref{theorem:KS} there is a kite $\kite_\alpha$ and two linear maps $\Psi : \R^3 \to V$, $\Phi : V \to \R^3$ such that $\Phi \circ \Psi = \Id$, $\Psi(\CC(\kite_\alpha)) \subset \C$ and $\Phi(\C) \subset \CC(\bsquare)$. Introduce now the extreme rays generators of $\CC(\kite_\alpha)$
\begin{equation*}
u_0=(-1,\alpha_0,1), \quad
u_1=(1,\alpha_1,1), \quad
u_-=(\alpha_-,-1,1), \quad
u_+=(\alpha_+,1,1).
\end{equation*}
Since the diagonals of a kite intersect, there exist 
positive numbers $\lambda_0$, $\lambda_1$, $\lambda_-$, $\lambda_+$ such that
\begin{equation*}
\lambda_0 u_0 + \lambda_1 u_1 = \lambda_- u_- + \lambda_+ u_+ .
\end{equation*}
Incidentally, those numbers can easily be expressed as elementary functions of the components of the vector $\alpha$; we do not report those formulae as we shall not need them in the following. We continue by defining the vectors
\begin{equation*}
x_i \coloneqq \lambda_i \Psi(u_i) \in \C, \quad i \in \left\{0,1,-,+\right\},
\end{equation*}
which by construction satisfy the identity
\begin{equation*}
    x_0+x_1=x_-+x_+.
\end{equation*}
On the other hand, let us consider the linear forms $T_i:\R^3\rightarrow \R$, $i \in \left\{0,1,-,+\right\}$, defined as
\begin{equation*}
T_0(x,y,z)\coloneqq z-x,\quad
T_1(x,y,z)\coloneqq z+x,\quad
T_-(x,y,z)\coloneqq z-y,\quad
T_+(x,y,z)\coloneqq z+y.
\end{equation*}
Define also the linear forms
\begin{equation*}
f_i \coloneqq T_i \circ \Phi: V \to \R, \quad i=\left\{0,1,-,+\right\}.
\end{equation*}
Using that 
\begin{equation*}
    \CC(\bsquare)  \subset  \left\{(x,y,z): z\geq 0,\, (x,y)\in [-z,z]^2 
    \right\},
\end{equation*}
we see that $T_i\in \CC(\bsquare)^*$ for every $i\in \left\{0,1,-,+\right\}$ and immediately infer that $f_i\in \C^*$. 
Moreover, it follows from the very definition of the $T_i$'s that
\begin{equation*}
f_0+f_1 = (T_0+T_1)\circ \Phi = (T_-+T_+)\circ \Phi = f_-+f_+.    
\end{equation*}
This proves claim~(i).

The identities in item~(ii) follow easily by plugging the concrete form of the $T_i$'s and $u_j$'s in the equation 
\begin{equation*}
f_i(x_j)=(T_i\circ \Phi)(\lambda_j \Psi(u_j))=\lambda_j\,T_i(u_j),
\end{equation*}
where in the last equality we have used that $\Phi \circ \Psi = \Id$.

Finally, let us verify item~(iii). Consider $0 \neq p\in \C$. Observe from Remark \ref{rem:ker-Phi} that $\Phi(p)$ is a nonzero element in $ \CC(\bsquare)$; writing $\Phi(p)=(x,y,z)$ with $z > 0$ and $(x,y)\in [-z,z]^2\setminus \{-z,z\}^2$, we see that
\begin{equation*}
(f_0+f_-)(p)=(T_0+T_-)(\Phi(p))=(T_0+T_-)(x,y,z)=2z-x-y>0.
\end{equation*}
The other cases can be analysed in an analogous fashion. This concludes the proof.
\end{proof}

Let us finally show how Theorem \ref{strong_incompatibility_thm} can be used to give a 
direct and operationally meaningful proof of Theorem~\ref{entangleability_thm}.

\begin{proof}[Proof (Theorem~\ref{entangleability_thm})]
The fact that if either $\C_1$ or $\C_2$ is a classical proper cone then $\C_1\tmin \C_2= \C_1\tmax \C_2$ is a well-known and easily verified fact (see for instance \cite[Lemma~5]{cones-1}).

In order to prove the converse, let us assume that $\C_1$ and $\C_2$ are both non-classical. Then, according to Theorem~\ref{strong_incompatibility_thm} there are non-zero vectors $0\neq x_0,x_1,x_-,x_+\in \C_1$, $0\neq y_0,y_1,y_-,y_+\in \C_2$ and functionals $f_0,f_1,f_-,f_+\in \C_1^*$, $g_0,g_1,g_-,g_+\in \C_2^*$ satisfying items (i)--(iii).

Let us construct the tensor 
\begin{align}\label{element w}
\omega\coloneqq x_0\otimes y_+- x_+\otimes y_++x_+\otimes y_0+x_1\otimes y_1.
\end{align}
One can easily check that $\omega \in \C_1\tmax \C_2$. Indeed, given $\varphi_1\in \C_1^*$ and $\varphi_2\in \C_2^*$, setting $a_i\coloneqq \varphi_1(x_i)$ and $b_i\coloneqq \varphi_2(y_i)$ for $i \in \{0,1,-,+\}$ gives that
\begin{align}\label{con max w}
(\varphi_1\otimes \varphi_2)(\omega)=a_0b_+-a_+b_++a_+b_0+a_1b_1=a_0b_0+a_1b_1-(a_+-a_0)(b_+-b_0)\geq 0.
\end{align}
Now, since $a_i\geq 0$ and $b_i\geq 0$ for every $i$, using item (i) from Theorem \ref{strong_incompatibility_thm} we obtain $-a_0\leq a_+-a_0= a_1-a_-\leq a_1$ and similarly for the $b_i$'s. This implies that $(a_+-a_0)(b_+-b_0)\leq \max\{a_0b_0,a_1b_1\}$, from which~\eqref{con max w} follows easily.

In order to finish the proof we will show that $\omega \notin \C_1\tmin \C_2$. To this end, let use define the linear functional 
\begin{align}
\varphi &= 2(f_0+f_1)\otimes (g_0+g_1) - (f_0-f_1)\otimes (g_0-g_1) - (f_0-f_1)\otimes (g_+-g_-) \\
&- (f_+-f_-)\otimes (g_0-g_1) + (f_+-f_-)\otimes (g_+-g_-).
\nonumber
\end{align}
The strategy for the rest of the proof is to show that $\varphi$ is strictly positive on $\C_1\tmin \C_2$ while $\varphi(\omega)\leq 0$. Indeed, let $\omega_1 \in \C_1$, $\omega_2 \in \C_2$, and define $c_1 \coloneqq (f_0+f_1)(\omega_1)$, $c_2 \coloneqq (g_0+g_1)(\omega_2)$, $d_1\coloneqq (f_0-f_1)(\omega_1)$, $d_2 \coloneqq (g_0-g_1)(\omega_2)$, $e_1 \coloneqq (f_+-f_-)(\omega_2)$, $e_2 \coloneqq (g_+-g_-)(\omega_2)$, we want to show that
\begin{equation*}
\varphi(\omega_1 \otimes \omega_2) = 2 c_1 c_2 - d_1 d_2 - d_1 e_2 - e_1 d_2 + e_1 e_2 >0.
\end{equation*}
Now, according to Theorem \ref{strong_incompatibility_thm} and Remark \ref{remark supp} right below it we know that $c_1=(f_0+f_1)(\omega_1)=(f_++f_-)(\omega_1)$, $c_2=(g_0+g_1)(\omega_2)=(g_++g_-)(\omega_2)$ are positive numbers; it then follows that $\tilde{d}_i\coloneqq d_i/c_i$, $\tilde{e_i} \coloneqq e_i/c_i$, for $i=1,2$, are all numbers in $[-1,1]$. Using item (iii) from Theorem~\ref{strong_incompatibility_thm}, we see that $(\tilde{d}_1,\tilde{e}_1)$ and $(\tilde{d}_2,\tilde{e}_2)$ both belong to $\bsquare$. Then, checking the inequality
\begin{equation*}
2 - \tilde{d}_1 \tilde{d}_2 - \tilde{d}_1 \tilde{e}_2 - \tilde{e}_1 \tilde{d}_2 + \tilde{e}_1 \tilde{e}_2 > 0
\end{equation*}
amounts to a straightforward computation.

It finally remains to show that indeed
\begin{equation*}
\varphi(\omega) \leq 0.
\end{equation*}
Through a long series of elementary algebraic manipulations one can verify that
\begin{equation}
\label{eq:magical}
\varphi(\omega) = 4\Big(f_0(x_+)-f_+(x_0)\Big)\Big(g_0(y_+)-g_+(y_0)\Big) .
\end{equation}
A SageMath script checking the above can be donwloaded at \url{https://github.com/gaubrun/entangleability}.

If the right-hand side of Eq.~\eqref{eq:magical} is nonpositive, we are done.
Otherwise, we could consider new elements $\tilde{\omega}$ and $\tilde{\varphi}$, defined as the previous ones by switching the role of $(x_0,x_1, f_0,f_1)$ and $(x_+,x_-, f_+,f_-)$, keeping untouched $y_i$ and $g_i$. These elements satisfy the same properties as $\omega$ and $\varphi$. Moreover, as one can observe from the right-hand side of \eqref{eq:magical}, we have $\tilde{\varphi}(\tilde{\omega}) = - \varphi(\omega) \leq 0$.
\end{proof}

\section{BB84 protocol in GPTs}

In this section we use Theorem~\ref{strong_incompatibility_thm} to construct a version of the BB84 protocol that works in any non-classical GPT. For a given non-classical GPT $(V,C,u)$, we can construct vectors $0\neq x_0,x_1;x_+,x_-\in C$ and functionals $f_0,f_1;f_+,f_-\in C^*$ satisfying conditions (i)--(iii) of Theorem~\ref{strong_incompatibility_thm}. Up to re-scaling, we can assume without loss of generality that $u(x_0+x_1)=1$, so that in fact $x_i=p_i\rho_i$, $x_j=q_j \sigma_j$ for some states $\rho_i,\sigma_j\in \Omega \coloneqq C \cap u^{-1}(1)$, with $p_i,q_j>0$ and also $p_0+p_1=q_++q_-=1$. Set $\omega\coloneqq p_0\rho_0+p_1\rho_1=q_+\sigma_++q_-\sigma_-$. Again, up to multiplying everything by a factor we can assume that $f_0+f_1 = f_++f_- \eqqcolon \ell \leq u$, so that $f_i,f_j$ are valid effects. Note that $\eta\coloneqq\ell(\omega)\geq \frac12 (f_0+f_+)(\omega) >0$ is a strictly positive constant.

Now, consider the following protocol to generate a secret key via a public noiseless channel connecting Alice to Bob and capable of transmitting states of the GPT $(V,C,u)$:
\begin{enumerate}[(1)]
\item Alice tosses $N$ times a fair coin. For each head, she prepares either $\rho_0$ (with probability $p_0$) or $\rho_1$ (with probability $p_1$); for each tail, she prepares either $\sigma_+$ (with probability $q_+$) or $\sigma_-$ (with probability $q_-$). She sends the states to Bob, in an orderly manner, using the channel $N$ times.
\item Bob tosses $N$ times a fair coin. If the $k^\text{th}$ coin is a head (respectively, a tail), he performs the measurement $\left(f_0,f_1,u-\ell\right)$ (respectively, $\left(f_+,f_-,u-\ell\right)$) on the $k^\text{th}$ state sent by Alice. 
\item Alice and Bob announce publicly the outcomes of their coins. They discard all rounds for which the outcomes of their coin tosses were different.
\item Also, Bob declares for which rounds he obtained the third measurement outcome, corresponding to $u-\ell$. Those rounds are also discarded.
\item If Eve has not interfered and the transmission were noiseless (something that can always be tested by sacrificing a small number of random bits), Alice and Bob are left with around $\frac12 N \ell(\omega)$ bits each. The bit strings held by Alice and Bob are equal, because the detection error probabilities are all zero, according to Theorem~\ref{strong_incompatibility_thm}(ii).
\end{enumerate}

The bits retained by Alice and Bob at the end of the above protocol are `flagged' by the corresponding outcomes of the coin tosses, and those outcomes, known to Eve, tell her something about the probability distribution of Alice and Bob's bits. Also, the fact that these bits have not been discarded in step~(4) skews the probability distribution of each bit as seen by Eve. For example, it is not difficult to verify that if the $k^\text{th}$ coin outcome was a head and the round has not been discarded, the probability that the encoded bit is $i\in \{0,1\}$ as seen by Eve is $p'_i = \frac{p_i\, \ell(\rho_i)}{\ell(\omega)}$. Importantly, $p'_i>0$, because if $0=\ell(\rho_i)=(f_0+f_1)(\rho_i))$ were to hold we would deduce that also $0=\ell(\rho_i) \geq f_j(\rho_i)$, so that $f_j(\rho_i)=0$; this would be in contradiction with the requirement that $f_i+g_i$ be strictly positive (Theorem~\ref{strong_incompatibility_thm}(iii)).

We can model the overall situation by saying that Alice, Bob, and Eve hold $n\approx \frac12 N \ell(\omega)$ i.i.d.\ random variable triples $(ZX,ZX,Z)$, where $Z\in\{0,1\}$ is the coin outcome and $X$ the bit encoded by Alice, so that
\bb
P_{ZX}(00)&=\frac{p'_0}{2}\, ,\quad P_{ZX}(01)=\frac{p'_1}{2}\, ,\\
P_{ZX}(10)&=\frac{q'_+}{2}\, ,\quad P_{ZX}(11)=\frac{q'_-}{2}\, ,
\ee
where according to the above reasoning
\bb
p'_i \coloneqq \frac{p_i\, \ell(\rho_i)}{\ell(\omega)}\, ,\qquad q'_j \coloneqq \frac{q_j\, \ell(\sigma_j)}{\ell(\omega)}\, .
\ee
In this situation, it is an easy corollary of the work of Maurer~\cite{Maurer1993} that the secret key rate $S(X\!:\!Y \| Z)$ Alice and Bob can achieve via public communication is given by~\cite[Theorems~2 and~3]{Maurer1993}
\bb
S(ZX:ZX \| Z) = H(X|Z) = \frac12 \left( h_2(p'_0) + h_2(q'_+) \right) > 0\, .
\label{secret_key_rate}
\ee
Although a fully-fledged proof of the security of the above protocol is beyond the scope of this work,~\eqref{secret_key_rate} indicates that our tweaked BB84 protocol achieves a non-zero secret key generation rate in every non-classical GPT.

\end{document}